# A Conceptual Framework for Definition of the Correlation Between Company Size Categories and the Proliferation of Business Information Systems in Hungary

PÉTER SASVÁRI, Ph.D.
ASSOCIATE PROFESSOR

e-mail: iitsasi@uni-miskolc.hu

SUMMARY

*Based on a conceptual model, this paper aims to explore the background of the decision-making process leading to the introduction of business information systems among enterprises in Hungary. Together with presenting the problems arising in the course of the implementation of such systems, their usage patterns are also investigated. A strong correlation is established between the size of an enterprise, the scope of its business activities and the range of the business information systems it applies.*
*Keywords: Business Information System, Internet, Hungary*
*Journal of Economic Literature (JEL) code: M15, M21, C10*

## INTRODUCTION

The role of information has become more and more substantial in the economy recently, and information is regarded as an important resource since it is more difficult for companies to improve their market positions in the long term without having the appropriate amount of available information (Cser and Németh 2007). Globalization in the business world has brought about the possibility of getting a greater amount of information in much less time, which means that companies are forced to spend more time and energy on handling the increased information load (Erdős 2005).

As business information systems are designed to provide effective help in this process, they are becoming increasingly popular among companies due to the robust technological development (Floyd and Wolf 2010). This paper deals with the usage of business information systems among Hungarian enterprises and analyzes the following three key questions: how the usage of business information systems influences a company's economic performance, what expenditure is required for an individual company to develop its information technology infrastructure and finally, to what extent information technology is considered important as a functional area within the organization of a company (Bubenko 2011).

The aim of the research presented in this paper was to explore the current situation of Hungarian enterprises in terms of using business information systems, gaining a more thorough insight into the background of the decisions made on introducing such information systems, together with the possible problems related to their introduction and further usage (Erdős 2005).

## THE CONCEPT OF BUSINESS INFORMATION SYSTEMS

There are several definitions offered on business information systems in the literature. According to Burt and Taylor's approach, "business information systems can be regarded as an information source in any combination thereof, or any access to and any recovery of their use or manipulation. Any business information system is designed to link the user to an appropriate source of information that the user actually needs, with the expectation that the user will be able to access the information satisfying their needs" (Burt and Taylor 2003: 52). Davis and Olson define business information systems as "an integrated user-machine system for providing information to support the operations, management, analysis, and decision-making functions in an organization. The system utilizes computer hardware and software, manual procedures, models for analysis, planning, control, and decision-making by using a database" (Davis and Olson 1985: 78).

"Information systems are a part of any organization that provides, generates, stores, separates, divides and uses information. They are made up of human, technical, financial and economic components and resources. In fact, they can be regarded as inherently human systems (organizations, manual systems) that may include a computer system, and automatizes certain well-defined parts and selected items of the system. Its aim is to support both the management functions and the daily operation of an organization." (Deák, Bodnár and Gyurkó 2008: 100)

In a broader sense, a business information system is the collection of individuals, activities and equipment employed to collect, process and store information related to the company's environment, its internal activities, together with all transactions between the company and its environment. Beyond giving direct support to operations, its basic task is to provide decision-makers with the necessary information during the whole decision-making process. The system's main components are the following (Drótos, Gast, Móricz and Vas 2006):

➢ Individuals carrying out corporate activities: the actual users of technical apparatus. Decision-makers also belong to this group, as leaders who receive information on the factors affecting business operations, and use business information systems to make decisions in relation to planning, implementation and monitoring business activities.

➢ Information (also known as processed data on external and internal facts) which – due to its systematized form – can be used directly in the decision-making process.

- Technical apparatus, nowadays usually a computer system that supports and connects the subsystems applied to achieve corporate objectives.
- The computer system standardizes a significant part of the information and communication system, thus making it easier to produce and use information.

According to one definition proposed (Csala et al., 2003: 110) "information systems are systems that use information technology to collect information, transmit, store, retrieve, process, display and transform information in a business organization by using information technology."

Raffai's understanding of information systems is as follows: "it uses data and information as a basic resource for different processing activities in order to provide useful information for performing useful organizational tasks. It's main purpose is the production of information, that is dedicated to creating messages that are new to the user, uncertainties persist, and their duties, to assist in fulfilling the decisions" (Raffai 2003: 60).

## ALTERNATIVES TO THE CLASSIFICATION OF BUSINESS INFORMATION SYSTEMS

The classification of business information systems is a difficult task because, due to the continuous development, it is hard to find a classification system that can present unanimously defined information system types. It occurs quite often that different abbreviations are used to refer to the same system or certain system types appear to be merged together. As a consequence, the classification of business information systems can be done in several ways, the lists of several groups of business information systems presented below just to show a few alternatives for classification (Bencsik 2011).

Dobay (1997) made a distinction between the following types:

- Office Automation Systems (OAS): used for efficient handling of personal and organizational data (text, image, number, voice), making calculations and document management.
- Communication systems: supporting the information flow between groups of people in a wide variety of forms.
- Transaction-processing systems (TPS): used for receiving the initiated signals of transactions, generating and giving feedback on the transaction event.
- Management Information Systems (MIS): used for transforming TPS-related data into information for controlling, management and analysis purposes.
- Executive Information Systems (EIS): intended to give well-structured, aggregated information for decision-making purposes.
- Decision support systems (DSS): applied to support decision-making processes with information, modelling tools and analytical methods.
- Facility Management Systems (facility management, production management): used for directly supporting the value production process.
- Group work systems: intended to give group access to data files, to facilitate structured workflows and the implementation of work schedules.

Another possible approach to defining categories is based on Raffai's work (2003):

- Implementation support systems: this group includes transaction processing systems (TPS), process control systems (PCS), online transaction processing systems (OLTP), office automation systems (OAS), group work support systems (GS), workflow management (WF), and customer relation management systems (CRM).
- Executive work support systems: this category can include strategic information systems (SIS), executive information systems (EIS), online analytical processing systems (OLAP), decision support systems (DSS), group decision support systems (GDSS), and management information systems (MIS).
- Other support systems: business support systems, (BIS), expert systems (ES), integrated information processing systems (IIS), and inter-organizational information systems (IOS) can be found in this category.

Based on Gábor's (2007) findings, business information systems can also be examined by applying the following classification criteria.

- According to organizational structure:
  - functional systems such as reporting applications,
  - comprehensive business systems such as corporate management systems used by the entire organization,
  - inter-organizational systems such as reservation systems.
- According to the field of application:

Depending on the scope of activities, systems used for accounting, finance, production, marketing or human resource management belong to this category. These systems are generally related to the various functions a company performs.

- According to the type of support:
  - TPS (Transaction Processing System) – it focuses on a particular purpose, its basic function is to serve as a supporting tool for data processing related to business activities.
  - MIS (Management Information System) – it basically supports functional executive activities (O'Brien 1999).
  - KMS (Knowledge Management System) – it facilitates the execution of tasks related to knowledge as a valuable corporate resource.
  - OAS (Office Automation System) – it supports office document management, group work and communication.
  - DSS (Decision Support System) – it supports decisions made by managers and analyses done by experts.
  - EIS (Enterprise Information System) – it is designed to support the whole organization and its management.
  - GSS (Group Support Systems) – it facilitates the cooperation between ad hoc and permanent work groups both within an organization and between different organizations.
  - ISS (Intelligent Support System) – it is mainly designed to support the work of employees performing mental work.
  - Applications supporting production activities: CAD/CAM (Computer Aided Design/Computer Aided Manufacturing) – they are designed to support planning and production processes by using information technology devices (Shaw 1991).

Finally, let us take a look at another classification system originally suggested (Kacsukné and Kiss 2007), which also served as a starting point for conducting the primary research presented in this article.

- TPS (Transaction Processing System): it is used for collecting, storing, modifying, and retrieving the daily transactions of a business organization. It usually consists of an advanced database system for such business events as settlement of accounts, sales, rental payments, orders and raw material purchases.
- MIS (Management Information System): it is used to analyze operational activities in the organization. It makes pre-defined reports at regular intervals even when special events occur; it focuses on the information need of managers and gives assistance to solve well-defined problems. It is efficient mainly at an operational or tactical level (Laudon 2009).
- DSS (Decision Support System): it is naturally emerged from management information systems, intended to help decision-makers to compile useful information from a combination of raw data, documents, and personal knowledge, or business models to identify and solve problems and make decisions. Its interactivity and the capability of elaborating problem-analysing models makes it especially effective at tactical levels.
- GDSS (Group Decision Support System): it is a further development of DSS where the stress is not at the level of personal decision-making; instead it supports joint decisions made by a group. Great emphasis is given to communication (e-mail, shared file access, video conferencing option).
- EIS (Executive Information System): it is designed to facilitate and support the information and decision-making needs of senior executives by providing easy access to both internal and external information relevant to achieving the strategic goals of a business organization. It is usually easy to use, offering user-friendly features.
- ERP (Enterprise Resource Planning System): its main purpose is to facilitate the flow of information between all business functions inside the boundaries of an organization and manage the relationships with outside stakeholders. It may include customer and supplier relationships and supply chain management as well. According to its most recent interpretation, it provides support to the full operational level by its modular structure.
- CRM (Customer Relationship Management System): it is designed to organize, automate, and synchronize business processes, mainly sales activities, but also those for marketing, customer service, and technical support. It also contributes to product development and the elaboration of marketing strategies.
- SRM (Supplier Relationship Management System): it is aimed at creating closer, more collaborative relationships with key suppliers in order to maximize the value realized through those interactions. As a cross-functional system, it provides support for decisions especially at operational and tactical levels (Hughes 2010).
- SCM (Supply Chain Management System): it is designed to facilitate the systematic and strategic coordination of the traditional business functions within a particular company and across businesses within the supply chain, for the purposes of improving the long-term performance of both individual companies and the supply chain as a whole. Its application is useful for making decisions both in operational and tactical levels (Harland 1996).
- BI (Business Intelligence System): it is designed to produce large amounts of information with the potential of leading to the development of new opportunities for a business organization. It often includes online analytical processing (OLAP), data mining, process mining, business performance management, benchmarking and predictive analytics. With its complexity, it proves to be one of the most powerful decision support tools.
- EPM (Enterprise Performance Management System): beside calculating performance indicators, its main task is to monitor and manage the hierarchy of indicators used to assess the overall performance of a business organization.
- KM (Knowledge Management System): it usually comprises a wide range of strategies and practices to identify, create, represent, distribute, and enable the adoption of knowledge. It is not strictly tied to managerial levels.
- ES (Expert System): it is designed to propose a solution to unstructured, specific problems where highly-prepared expertise is needed. It actually stores all the available facts and figures, then it draws conclusions based on them. Actually, the facts and rules are stored, and based on these conclusions. It is a special field of application within the broader area of artificial intelligence.

## FACTORS AFFECTING THE IMPLEMENTATION OF BUSINESS INFORMATION SYSTEMS

When a business organization makes a decision about introducing any business information system, their decision can be explained by a variety of factors. The most common factors are as follows (Kacsukné and Kiss 2007: 245):

- "Technical considerations: companies applying fragmented, outdated business information systems with the lack of transparency.
- Strategic considerations: ERP systems may play a role in maintaining and enhancing competitiveness, they may establish the technical background to apply e-commerce solutions.
- Business considerations: among others, cost reduction and profit increase objectives, job cuts, stock reduction, reducing IT costs, improving productivity and more rapid turnaround of orders may belong to this group of factors."

Ideally, before a company decides to introduce a business information system, they consider a large number of factors. The most important step during this process is to select the most relevant aspects, then, after weighing them carefully, the management of a company can choose the best offer available. According to Kacsukné and Kiss (2007), these aspects can be the following:

General aspects

- Availability of documentation: it is also important for a company to investigate the availability of user guides, manuals and other system support documents.
- Compatibility: it is also an important matter if the newly introduced system fits the existing hardware and software assets, and whether it is compatible with the hardware and software devices available on the market.
- Costs: in the process of introducing a business information system, a business organization not only has to pay the price of a software product but it also has to pay attention to the additional costs related to its introduction such as education, professional and license fees, not to mention some incurring costs during its usage (telecommunications, maintenance and repair costs). In order to make an optimal decision, it is recommended to consider some other indirect effects of the introduction as well.

➣ Ergonomy: the user-friendly nature of a system or an application is monitored here, with a special emphasis on their effects on the human nervous system, the eyes and hands.
➣ Modularity, expendability: in the market of business applications, companies generally purchase the modules of various business information systems that are necessary to perform certain functions, maintaining the possibility of adding more modules to the purchased system in the future.
➣ Network access: it also should be considered whether the newly-implemented system can be integrated into the existing network. In the case of hardware, it is a question of physical interface, whereas in the case of software the real question to be considered whether the new application can run in a network environment.
➣ Performance: a decision can be deeply affected by such information on the performance of the desired business information system as speed, capacity features, and the necessary operating systems for its usage.
➣ Reliability: it is important to determine whether there is a risk of failure and the extent to which occasional errors may result in damage. There are available systems that already have built-in self-monitoring and error diagnostic functions. The criterion of reliability is particularly important in those areas where human life is at stake or the occurrence of a failure may end up in causing huge financial losses (in hospitals, air traffic control, banks, etc.).
➣ Support service: it may also be an important factor to what extent the manufacturer provides the installation, maintenance and repair of the newly-introduced system.
➣ Technology: as in the case of products, product life cycle is a crucial factor in business information systems. A business organization has to decide whether to take the risk of experimenting with a brand-new technology or to resort to using more proven but less modern systems.
➣ The manufacturer's reputation: although this aspect is not included in the referenced literature, it is possible that some companies prefer to ask for an offer from a larger, more respected service provider, ignoring smaller companies that may provide the same services with the same quality.
➣ Usability: first, a business organization has to consider whether the applicable system is suitable for the tasks it is required to perform. If it turns out that the selected system is only partially able to fulfill the requirements, decision-makers will have to make compromises in terms of their needs, after taking other aspects into account.
➣ Warranty: this includes the evaluation of the warranty services and conditions provided by the manufacturer.

Specific aspects
➣ Availability of new software versions: it is reasonable to think about the future when selecting a business information system, that is, to check if there will be any new versions available for the selected system, what areas will be affected by the occasional upgrades, how quickly they will be done, and what additional costs will be incurred.
➣ Customer support: this means that a business organization needs to know whether the producer is willing to provide customer support in the phase of introduction and after introduction if needed.
➣ Flexibility and customization: how much flexibility is provided to serve unique customer needs in connection with the construction of the system may depend on the type of the applied system or on the developer company. It is therefore important to take into account to what extent the implementation of a new system can be adapted to already established business processes. It is not reasonable to change well-functioning and long-established business processes in order to meet the capabilities of a newly introduced system, just because the new system is unable to adapt to the specific needs of the company.
➣ Free trial period: it is possible that some companies may choose a system based on experiences gained during a product demonstration or the testing of a shareware application.
➣ Security: it is very important for a company to take into account protection options against the possibility of causing intentional or accidental damage to a company's existing network system.

## THE AIM AND THE CONCEPT OF THE RESEARCH

The review of the relevant literature on the subject made it possible to identify the most important points of the research. Based on these, the main objectives as well as the concept of the research were formulated. The research objectives are the following:
➣ to present the background of the decisions related to the introduction of business information systems, along with the problems encountered in the phase of their introduction,
➣ to analyze the usage patterns of business information systems,
➣ to reveal the connection between using business information systems and the operational effectiveness or profitability of companies.

Based on the aims presented above, the following research concept was determined:
➣ First, the major issues related to the introduction of business information systems were analyzed. It was surveyed whether the companies taking part in the study used any sort of business information systems, and if not, what causes or conditions prevented them from introducing them. In the case of companies applying business information systems, the causes of introducing such systems, the information sources for selecting the appropriate systems, and the criteria for selecting them were also investigated. It was also examined whether companies had made calculations on the costs of introducing business information systems before making decisions on them, and if so, what aspects had been taken into account during the calculation. The problems occurring in the phase of implementation were identified.
➣ After that, the usage patterns of business information systems at companies were examined. The main points of the relevant analysis were the given company's information technology infrastructure, its Internet usage habits, and its appearance on the Internet. Here, the types of the applied business information systems and their areas of use were also presented, then the forms of information technology functions, human resource issues and the main points of IT strategy were covered.
➣ In the closing part of the analysis, the impacts of using business information systems on the operational effectiveness of companies were examined. It was investigated whether the introduction of such systems had an influence on the performance and revenues of the company as well as the size of the targeted market and the changes on the demand side. Another point of the investigation was whether ensuring more efficient information flow and information management contributed to the reduction of the company's other costs. These factors, summarized in Figure 1, of course, cannot be quantified so easily; however, taking them into consideration can lead to making more firmly grounded decisions on the introduction of various business information systems.

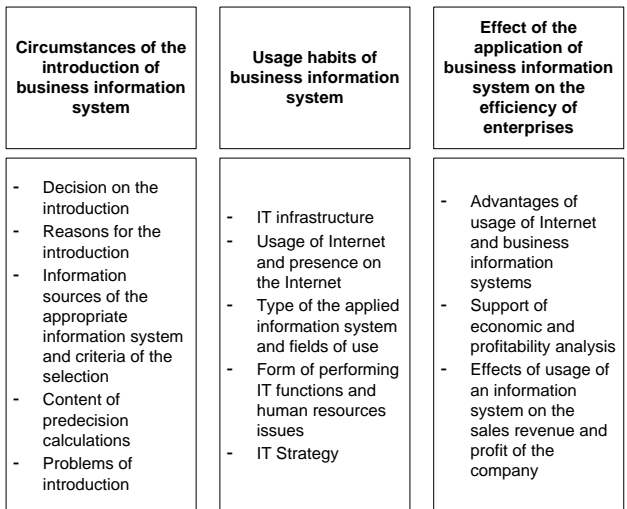

Source: own elaboration

*Figure 1. Conceptual model of the primary research*

## THE RESEARCH METHOD

The empirical survey was carried out using a written questionnaire. In the phase of compiling the individual questions of the survey, the main results of the previously conducted empirical surveys on the subject were also taken into account.

The questionnaire was divided into five major parts. The first part included some basic questions about the companies' background (such as their location, fields of operation, number of employees etc.), then questions related to the responding company's information technology infrastructure followed. In the third part of the questionnaire, the emphasis was put on questions enquiring about the Internet-using habits of the companies; the fourth group of questions was aimed at enquiring about the usage patterns of business information systems, making it the most detailed part of the questionnaire. The closing part contained questions about the IT-skilled human resources employed by the responding companies.

The questionnaire was sent out to several hundreds of companies, The Hungarian survey was conducted both in a paper-based format and online with the assistance of the software application called Evasys. For evaluating data and presenting the results of the survey, the statistical software packages Excel 2007 and SPSS 19.0 were applied.

## THE BACKGROUND CONDITIONS OF INTRODUCING BUSINESS INFORMATION SYSTEMS OF THE SURVEYED COMPANIES

The circumstances of decisions concerning the introduction of business information systems were analyzed. The first task was to gain more information on whether the surveyed companies apply any kind of business information system.

Based on the results, it can be stated that 17% of the respondents do not apply any kind of business information system and do not plan their introduction, while 11.7% of them were not applying any kind of business information systems at the time of being questioned but did not exclude the possibility of introducing such systems later.

It was interesting to examine the reasons why a company did not use any of above-described business information systems. Mostly the size of the company justifies why a company does not apply any kind of business information system. More than three-quarters of the microenterprises do not think about using business information systems due to the size of the company. In addition, the financial resources of the company may influence the decision on the introduction of a business information system; a quarter of the microenterprises stated that the lack of financial means was behind ignoring such systems. In some other cases, business information systems were not introduced due to the lack of management need; however, this is only typical of small and medium-sized enterprises. A statistical analysis is shown in Table 1, demonstrating that in many cases the lack of senior management interest prevented the introduction of a new business information system. However, no significant correlation was found between company size and the possible causes of not introducing any business information systems.

*Table 1*
*Reasons for the decision not to use a business information system*
*(Phi, Cramer's V and Contingency Coefficient values)*

| Possible reasons of the decision | Phi | | Cramer's V | | Contingency Coefficient | |
|---|---|---|---|---|---|---|
| | Value | Approximate Significance | Value | Approximate Significance | Value | Approximate Significance |
| The lack of senior management interest prevented the introduction of such information systems* | 0.213 | 0.233 | 0.213 | 0.233 | 0.209 | 0.233 |
| The introduction of such information systems is not required because of the size of the company | 0.600 | 0.000 | 0.600 | 0.000 | 0.514 | 0.000 |
| The company's financial means do not allow the introduction of such information systems* | 0.239 | 0.145 | 0.239 | 0.145 | 0.233 | 0.145 |

* No correlation
Source: own elaboration

Companies were asked to rank the selection criteria for introducing business information systems on a scale of 5. As can be seen in Figure 2, usability proved to be the most important factor, that is, that the selected information system would be capable of performing the necessary tasks. This was regarded as the most vital criterion mostly by microenterprises.

The value of reliability as a criterion reached 4.5 on average, which shows that companies attribute great importance to how high the risk of system failure can be.

| | Micro-enterprise | Small-sized enterprise | Medium-sized enterprise | Corporation |
|---|---|---|---|---|
| **General aspects:** | | | | |
| Usability | ✔ 5,00 | ✔ 4,46 | ✔ 4,60 | ✔ 4,83 |
| Reliability | ⚠ 4,13 | ✔ 4,46 | ✔ 4,60 | 4,56 |
| Network access | ✔ 4,38 | ⚠ 4,23 | ✔ 4,50 | 4,67 |
| Compatibility | ⚠ 3,75 | ✔ 4,54 | ✔ 4,38 | ✔ 4,56 |
| Performance | ⚠ 3,88 | ✔ 4,38 | ✔ 4,32 | ✔ 4,28 |
| Costs | ✔ 4,63 | ⚠ 4,23 | ⚠ 4,12 | 4,17 |
| Support service | ⚠ 3,75 | ⚠ 3,92 | ✔ 4,40 | 4,33 |
| Modularity, extendability | ⚠ 4,13 | ⚠ 4,08 | ⚠ 4,04 | ✔ 4,50 |
| Warranty | ⚠ 3,63 | ⚠ 3,69 | ⚠ 4,20 | 4,22 |
| Technology | ✖ 3,25 | ⚠ 3,62 | ⚠ 4,00 | 4,17 |
| Availability of documentation | ⚠ 3,50 | ⚠ 3,92 | ⚠ 3,92 | 4,00 |
| Ergonomy | ✖ 3,25 | ⚠ 3,69 | ⚠ 3,58 | 3,59 |
| Manufacturer's reputation | ✖ 2,75 | ✖ 3,31 | ✖ 3,29 | ✖ 3,17 |
| **Specific aspects:** | | | | |
| Security | ✔ 4,38 | ✔ 4,31 | ✔ 4,40 | ✔ 4,39 |
| Flexibility, customization | ✔ 4,38 | ⚠ 4,15 | ✔ 4,28 | 4,17 |
| Customer support in the phase of introduction | ✖ 2,88 | ⚠ 4,08 | ⚠ 4,16 | 4,06 |
| Compliance with information strategy | ⚠ 3,63 | ⚠ 3,77 | ⚠ 4,04 | 3,94 |
| Availability of new software versions | ⚠ 3,63 | ✖ 3,42 | ⚠ 3,92 | 3,89 |
| Free trial period | ✔ 4,38 | ⚠ 3,85 | ⚠ 3,64 | 3,67 |
| Customer support after introduction | ✖ 2,86 | ⚠ 3,67 | ⚠ 4,04 | 4,00 |

check mark ✔ = higher than average, ✖ = average, ⚠ = lower than average

Source: own elaboration

*Figure 2. Selection criteria for applying business information systems: average points by company size*

Several criteria reached a higher average value than 4. According to the scores, the main criteria to be considered are whether the system operates in a network environment, fits in the existing system environment, and secures protection against accidental or intentional damage. In addition, companies consider the performance, the costs and the customization of the system as well. The consideration of costs may receive greater emphasis in case of microenterprises, because fewer resources are at their disposal to implement a business information system compared to the opportunities of a corporation. The companies also marked the assurance of service and support, future expansion possibilities of the system and the warranty conditions as important criteria. It can be seen that there are significant differences in the evaluation of companies by company size regarding the above criteria. The responding microenterprises gave significantly lower values the customer support both during and after introduction. One possible reason for this can be that introducing business information systems for customer support requires more resources from the company; the lack of capacity may prevent them from introduction. Compatibility as a criterion is also was also given lower values in their responses, which can be due to the fact that it rarely occurs when the new system fits into a previously introduced system, either because the new one is not compatible with existing softwares or its complex features do not meet the basic needs of microenterprises.

Another important goal of the research is to find an answer to the question of whether the enterprises carried out calculations on expenditures when making a decision on the introduction of a business information system and the selection of the appropriate system, and what items were taken into account during the calculation. The question was justified by the often-raised issue that following the introduction of a system, numerous hidden costs came to light on which the companies had not counted.

Apart from taking into account expenditures, it was also relevant to ask whether the companies analyzed the possible effects of the introduction of business information systems on sales revenue, profits and other costs.

The following expenditure items related to business information systems were taken into consideration: investment and development expenses, repair, maintenance, cost of spare parts, telecommunications expenses, costs corresponding to education and training, professional fees, licence fees and personnel expenses.

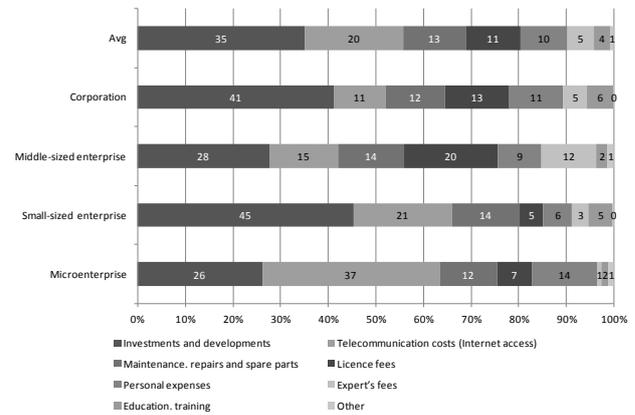

Source: own elaboration

*Figure 3. The structure of actual annual IT expenses reported, average values by company size*

Table 2 contains important information about several aspects concerning the content of calculations prior to the introduction of a business information system. Several factors were listed in the questionnaire concerning the content of the calculation. The respondents were required to declare whether they had taken into account the given factors during their calculations or not.

*Table 2*
*IT expenses and economic effects taken into account before the introduction of business information systems*

| IT-expenses and economic effects | Corporation | Medium-sized enterprise | Small-sized enterprise | Micro-enterprise |
|---|---|---|---|---|
| Investment and development costs | 77.78% | 76.47% | 88.46% | 100.00% |
| Personal expenses | 88.88% | 76.47% | 69.23% | 100.00% |
| Telecommunication costs | 77.78% | 93.75% | 65.38% | 89.48% |
| Cost of maintenance and repairs | 55.55% | 76.47% | 84.00% | 100.00% |
| Licence fees | 33.33% | 88.24% | 92.31% | 94.74% |
| The indirect effect of introducing the new information system on revenues | 55.56% | 78.57% | 69.23% | 84.21% |
| The indirect effect of introducing the new information system on profits | 55.55% | 64.71% | 61.53% | 83.33% |
| Expert's fees | 33.33% | 64.70% | 76.92% | 84.21% |
| Costs related to training and education | 33.33% | 58.82% | 60.00% | 84.21% |
| The indirect effect of introducing the new information system on demand | 22.22% | 64.71% | 68.00% | 66.67% |
| The indirect effect of introducing the new information system on other costs not related to information technology | 44.44% | 21.43% | 60.00% | 73.68% |

Source: own elaboration

Based on these findings, the following connections came to light. Ranking of certain IT expense items within the total expenditures is also reflected in whether companies took these expenses into account during the calculations carried out before the introduction. It can be seen in Table 2. That, based on the total data of the responses, investment and development expenses, telecommunication costs, personnel expenses as well as repair and maintenance costs were mainly considered during the calculations. Compared to the relatively low rate of corporations, the companies of other size categories gave great importance to licence fees in their calculations.

Some significant correlations arose in terms of size categories as well. The average values of companies belonging to the largest size category proved to be the highest almost in all aspects. As a consequence, corporations carry out the most extensive and careful calculations before the introduction of a business information system. In their case, both personnel and expertise conditions were available for performing such calculations. On the contrary, in the case of microenterprises, the lowest values were shown in the majority of the factors. For microenterprises, as they have fewer personnel in order to carry out such economic examinations, although raising funds for investments and operating resources means the greatest problem for these companies. Strong correlations with the size of the company can also be discovered in the cases of several other items.

## THE USAGE PATTERNS OF BUSINESS INFORMATION SYSTEMS

In connection to the IT infrastructure, the following two questions were to be answered: whether a server-based network operated at the company, and the total number of computers operating at the company site. Based on the received figures, it can be asserted that two-thirds of the respondents (69.1%) operate a server-based network. Considering the number of the computers at the company, correspondence with the size of the companies is natural. At corporations the average number of the computers was 549, at medium-sized enterprises there were 55, at small-sized companies there were 7, at microenterprises there were only 3 computers on average.

Perhaps it is not surprising in today's world that all of the respondent companies have Internet access. Among the objectives of using the Internet, there are a few remarkable differences by size categories. As Figure 4 shows, primarily corporations use the Internet for education purposes. Besides the use of tax advisory services and purchase of goods and services there are no big differences between the purpose of use according to size categories, however it can also be realized that in most categories the ratio of corporations are lower compared to other size categories.

In terms of "other" purposes of use, several responses were received that could not be classified into the optional categories, for example submission of project tender applications, website updates, benchmarking, the use of a web-based trading system, access to the central database via company programmes, service providing via the Internet, connection to external partners and companies through a part of the company's network, development of new services, development of new services related to education, seeking long-term business partners.

Of the responding companies, 86.2% have a website. This figure also shows that nowadays a website is already a standard tool for the majority of the companies and a presence on the Internet is becoming more and more natural.

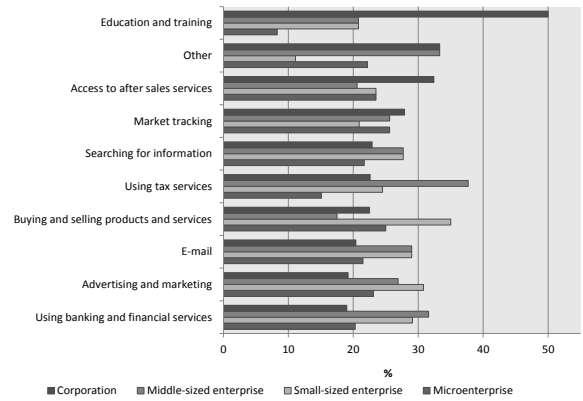

Source: own elaboration

*Figure 4. The purpose of using the Internet among the surveyed companies by size category*

Every company website offers a wide range of information and services. Companies having a website provided the services listed in Figure 5. Not surprisingly, most of the information placed on their websites is connected to the companies and the products and services they offer. In terms of company size, mainly medium-sized companies and corporations use their websites for this purpose. In addition, the most common features are providing customer service such as e-mail or a forum for their products and services, sales of products and services, placing job advertisements, and receiving online orders. In order to carry out secure transactions or provide online digital services and online payment options, a much more complex website is required, whose maintenance and development needs major resources. This could explain the fact that these options are provided only by medium-sized companies and corporations.

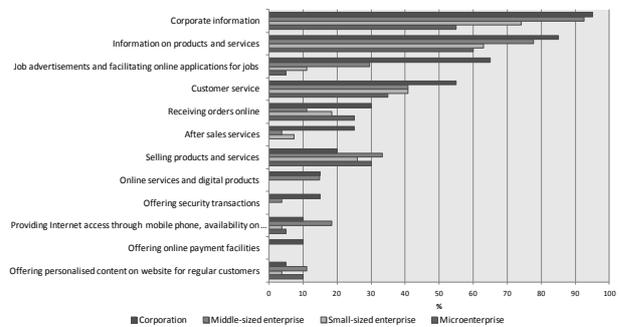

Source: own elaboration

*Figure 5. On-line services provided by companies based on their size*

Business information systems were applied in different areas, representing a different development level listed in the questionnaire. The respondents had to declare if there was an operating business information system of the kind at their company; if the answer was no, they were asked whether they were planning to introduce such a system later.

Three-quarters of the corporations use transaction processing systems (TPS), and one-tenth of them are planning to introduce TPS in the future. More than half of the medium-sized companies and almost a third of small-sized enterprises also use such systems; however, the number of microenterprises is irrelevant in this regard.

Similar ratios could be detected in the case of office automation systems (OAS) and enterprise resource planning systems (ERP), with the only difference that the latter could not be found in microenterprises and only 10% of them were planning to start applying such systems.

Supplier relationship management systems (SRM) are used by nearly two-thirds of corporations, while the same rate among medium-sized enterprises is only 37%. The use of these systems by microenterprises was insignificant.

Supply chain management system (SCM) applications are used by 40% of corporations and the remaining ones do not plan their introduction. A fourth of medium-sized companies already use supply chain management systems and another fourth of them are planning their introduction. About one-tenth of small-sized businesses apply such systems and there is a very small proportion of microenterprises using them.

Half of the corporations and nearly half of the medium-sized companies have customer relationship management systems (CRM) in use. More than a third of microenterprises are planning to introduce CRM systems in the future but their scale still remains very small. Geographic information systems (GIS) are used primarily by corporations, with a relatively high proportion of 40%, but surprisingly, some microenterprises also operate geographic information systems and a further 15% of them are planning to apply GIS in the near future.

An Intranet operates at the vast majority (reaching 80%) of corporations, more than one-third of the medium-sized companies also have internal network, and in addition, at the small and micro-enterprises it is operated or it is planned to be implemented in a similar proportion.

It was also investigated whether there was a relationship between company size and the use of information systems in different operational areas of the company. It was proved by the help of a cross-analysis that there was a significant relationship between fifteen operational areas and company size (Table 3).

*Table 3*
*Use of information systems according to the operational areas of enterprises (Phi, Cramer's V and Contingency Coefficient values)*

| Operational areas | Relationship | Phi | | Cramer's V | | Contingency Coefficient | |
|---|---|---|---|---|---|---|---|
| | | Value | Approximate Significance | Value | Approximate Significance | Value | Approximate Significance |
| Accounting | moderate | 0.540 | 0.000 | 0.540 | 0.000 | 0.475 | 0.000 |
| Finance | moderate | 0.565 | 0.000 | 0.565 | 0.000 | 0.492 | 0.000 |
| Salary and wage administration | moderate | 0.648 | 0.000 | 0.648 | 0.000 | 0.544 | 0.000 |
| Human resource management | weaker than moderate | 0.413 | 0.000 | 0.413 | 0.000 | 0.382 | 0.000 |
| Senior management decision support | - | 0.350 | 0.009 | 0.350 | 0.009 | 0.330 | 0.009 |
| Controlling, planning | moderate | 0.547 | 0.000 | 0.547 | 0.000 | 0.480 | 0.000 |
| Purchasing | weaker than moderate | 0.458 | 0.000 | 0.458 | 0.000 | 0.417 | 0.000 |
| Stockpile management | weaker than moderate | 0.492 | 0.000 | 0.492 | 0.000 | 0.441 | 0.000 |
| Asset management | moderate | 0.514 | 0.000 | 0.514 | 0.000 | 0.476 | 0.000 |
| Maintenance | weaker than moderate | 0.483 | 0.000 | 0.483 | 0.000 | 0.435 | 0.000 |
| Production/services | - | 0.365 | 0.006 | 0.365 | 0.006 | 0.343 | 0.006 |
| Sales, invoicing | moderate | 0.525 | 0.000 | 0.525 | 0.000 | 0.465 | 0.000 |
| Environmental management | - | 0.220 | 0.208 | 0.220 | 0.208 | 0.215 | 0.208 |
| Customer service | - | 0.291 | 0.047 | 0.291 | 0.047 | 0.279 | 0.047 |
| Marketing | weak | 0.367 | 0.005 | 0.367 | 0.005 | 0.345 | 0.005 |
| Administration | - | 0.341 | 0.012 | 0.341 | 0.012 | 0.322 | 0.012 |
| Quality control | weak | 0.389 | 0.003 | 0.389 | 0.003 | 0.363 | 0.003 |
| Project management | - | 0.087 | 0.869 | 0.087 | 0.869 | 0.087 | 0.869 |

Source: own elaboration

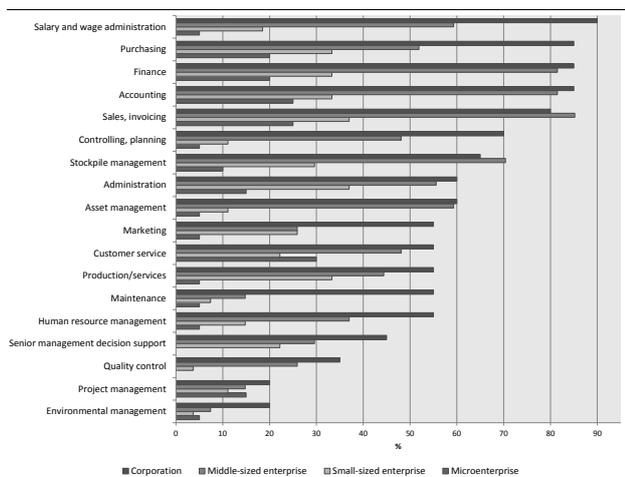

Source: own elaboration

*Figure 6. Use of information systems according to operational areas by company size*

As can be seen in Table 3, there is a moderate relationship in the fields of salary and wage administration, accounting, finance, controlling, planning, tangible asset management, sales and invoicing, while a weaker-than-moderate relationship can be observed in the fields of human resources, maintenance, purchasing, stockpile management, administration, production, service and management support. A weak relationship was detected in the fields of marketing and quality assurance.

CONCLUSIONS

Nowadays the issue of information technology in business is moving into the centre of attention, which is also indicated by the fact that more and more companies, not accidentally, recognize its importance. Business information systems are not only fashionable – their application promotes more efficient operation of the company and also improves the supply of information to decision-makers; applying such systems can also play an important role in helping companies to put greater emphasis on information technology in order to gain a competitive advantage.

My aim was to present the circumstances of the decisions made about the introduction of business information systems and problems emerging during the introduction as well as to analyze the usage habits of companies applying these systems, and to explore the relation between the application of business information systems and the operational effectiveness of the business.

Based on the scientific literature, I worked out a conceptual model appropriate for the aims of the research, serving as a base both for the questionnaire and the analysis. The primary focus of the analysis was to explore the differences and similarities of the usage habits of business information system by size categories. Thus, the micro-, small and medium-sized enterprises as well as corporations were also presented in the sample.

According to my observation, the correlation between the given factors could even be further strengthened by the application of complex statistical methods and by performing additional correlation assessments where the comparison should be carried out based on the main activity of the company rather than the size of the company, as I assume that the business scope of a company also determines the range of business information systems in use.

*Acknowledgements*

*The described work was carried out as part of the TÁMOP-4.2.2/B-10/1-2010-0008 project in the framework of the New Hungarian Development Plan. The realization of this project is supported by the European Union, co-financed by the European Social Fund.*